\documentclass[aps,prl,twocolumn,superscriptaddress,floatfix,showpacs]{revtex4}
\usepackage[dvipdfmx]{graphicx}
\usepackage[dvipdfmx]{color}
\usepackage{amsmath}
\usepackage{amssymb}
\usepackage{dcolumn}
\usepackage{bm}
\usepackage{latexsym} 
\usepackage{setspace}
\usepackage{graphicx}
\usepackage{xcolor}
\usepackage{ulem}
\usepackage{braket}
\begin{document}
\title{Wannier-based implementation of the coherent potential approximation with applications to Fe-based transition-metal alloys}
\author{Naohiro Ito}
\email{nao.ito@cmpt.phys.tohoku.ac.jp}
\affiliation{Department of Physics, Tohoku University, Sendai 980-8578, Japan}
\author{Takuya Nomoto}
\affiliation{Department of Applied Physics, The University of Tokyo, Hongo, Bunkyo-ku, Tokyo 113-8656, Japan}
\author{Koji Kobayashi}
\affiliation{Institute for Materials Research, Tohoku University, Sendai 980-8577, Japan}
\author{Sergiy Mankovsky}
\affiliation{Department Chemie, Physikalische Chemie, Universit\"at M\"unchen, Butenandstr. 5-13, 81377 M\"unchen, Germany}
\author{Kentaro Nomura}
\affiliation{Institute for Materials Research, Tohoku University, Sendai 980-8577, Japan}
\affiliation{Center for Spintronics Research Network, Tohoku University, Sendai, Miyagi 980-8577, Japan}
\author{Ryotaro Arita}
\affiliation{Department of Applied Physics, The University of Tokyo, Hongo, Bunkyo-ku, Tokyo 113-8656, Japan}
\affiliation{RIKEN Center for Emergent Matter Science (CEMS), Wako 351-0198, Japan}
\author{Hubert Ebert}
\affiliation{Department Chemie, Physikalische Chemie, Universit\"at M\"unchen, Butenandstr. 5-13, 81377 M\"unchen, Germany}
\author{Takashi Koretsune}
\affiliation{Department of Physics, Tohoku University, Sendai 980-8578, Japan}
\begin{abstract}
We develop a formulation of the coherent potential approximation (CPA) on the basis of the Wannier representation to develop a computationally efficient method for the treatment of homogeneous random alloys that is independent on the applied first-principles electric structure code.
To verify the performance of this CPA implementation within the Wannier representation, we examine the Bloch spectral function, the density of states (DOS), and the magnetic moment in Fe-based transition-metal alloys Fe-$X$ ($X$ = V, Co, Ni, and Cu), and compare the results with those of the well-established CPA implementation based on the KKR Green's function method.
The Wannier-CPA and the KKR-CPA lead to results very close to each other.
The presented Wannier-CPA method has a wide potential applicability to other physical quantities and large compound systems because of its low computational effort required. 
\end{abstract}
\maketitle
\section{I. Introduction}
Many substitutional alloys show a fascinating richness in their physical properties depending on their composition.
For example, in spintronics, the spin Hall angle can be tuned by alloying \cite{Obstbaum2016PRL}.
Another example is the possibility to induce magnetism in semiconductors by the addition of impurities \cite{Furdyna1988JApplPhys}.
\par
The methods to calculate the electronic structure of substitutional alloy systems have been developed since the 1930s \cite{Faulkner1982ProgMaterSci,Temmerman1987CompPhysRep,Economou2006Springer}.
The simplest approach for calculations on alloys is the virtual crystal approximation (VCA), in which the concentration average of the potential is placed on each site of the lattice \cite{Nordheim1931AnnPhysik,Muto1938SciPapers}.
Although the VCA seems to be a good approximation for metals having a simple free-electron-like electronic structure such as Na, K, and Al, it is known that the VCA completely fails to yield correct physical properties for the transition-metal alloys \cite{Temmerman1987CompPhysRep}.
In particular, the VCA fails to describe element-specific properties of an alloy as it is relevant for example in hyperfine interaction \cite{Faulkner1982ProgMaterSci}.
This shortcoming of the VCA has been removed by Korringa \cite{Korringa1958JPhysChemSolids} and Beeby \cite{Beeby1964PhysRev} by introducing the so-called average $t$-matrix approximation (ATA).
Within this approach, the concentration average of the single-site scattering matrix, the $t$-matrix, is used instead of the potential to consider component-projected properties.
However, the ATA still has formal problems leading sometimes to unphysical results \cite{Faulkner1982ProgMaterSci}.
For homogeneous random alloys, the most sophisticated single-site method solving these problems is provided by the coherent potential approximation (CPA).
The CPA is a mean-filed theory treating alloys by introducing an effective medium defined by its average scattering properties first proposed by Soven \cite{Soven1967PhysRev} and Taylor \cite{Taylor1968PhysRev}.
Formulating the CPA within the framework of multiple scattering or Korringa-Kohn-Rostoker (KKR) formalism implies that embedding one of the alloy components into the CPA should lead to no additional scattering on average.
Accordingly, unlike for the VCA and ATA methods, one has to determine the effective medium self-consistently for the CPA calculations.
\par
As the CPA can be easily applied on the basis of electronic structure methods working with the Green's function, it is usually formulated by a combination with the tight-binding (TB) method or the KKR Green's function method \cite{Shiba1971ProgTheorPhys,Gyorffy1974JPhysColloq}, which is a well-established first-principles electronic structure calculation method.
Especially using the KKR-CPA method \cite{Gyorffy1972PRB,Faulkner1980PRB} quite a number of physical properties of alloy systems have been studied, such as their magnetic structure properties of dilute magnetic semiconductors \cite{Akai1998PRL}, exchange coupling, and the corresponding magnetic transition temperature \cite{Sato2005HyperfineInteract,Ebert2009PRB}, as well as transport properties as the extrinsic and intrinsic contributions of the anomalous \cite{Lowitzer2010PRL} and the spin Hall effect \cite{Lowitzer2011PRL}.
This situation is due to a characteristic feature of the KKR method.
Unlike other general first-principles calculation methods, such as the standard pseudopotential-based methods or the linearized augmented plane wave (LAPW) method, the Green's function of the system is used already within the self-consistent field (SCF) calculation step when performing KKR calculations.
Therefore, it is easy to construct the Green's function for alloy systems by means of the KKR-CPA and directly calculate physical quantities using the resulting Green's function.
In the field of first-principles calculations, the CPA is alternatively formulated on the basis of the tight-binding linear muffin-tin orbital (TB-LMTO) method \cite{Kudrnovsky1987PRB} as well as the linear combination of atomic orbitals (LCAO) methods \cite{Richter1988JPhysFMetPhys,Richter1987JPhysFMetPhys}.
\par
In this contribution, we present an implementation of the CPA similar to the KKR-CPA Green's function method but more efficient and widely applicable while keeping the accuracy for the prediction of physical properties that the KKR-CPA method possesses.
For the practical realization of this goal, we focus on the Wannier formulation.
The reason for this is that we can construct Wannier functions from any kind of first-principles calculation method if the wave function is available, and set up a corresponding TB model from the obtained Wannier functions.
This means that we present a computational method for the electronic structure of random alloys that can be combined with any kind of first-principles computational method.
Moreover, we can substantially reduce the computational time when using the Wannier formalism if it is successfully combined with the CPA as it can be performed independently from the SCF calculations done by the first-principles calculations.
The Wannier functions are widely used to investigate physical properties \cite{Marzari2012RevModPhys,Nomoto2020PRB}.
However, only very few studies were done for alloy systems \cite{Mourada2012EurPhysJB,Koehl2012OptLett}.
One of the reasons for this is that there is an ambiguity in the determination of the relative reference values of the on-site potentials for the elements that form an alloy.
Concerning this problem, we propose a very simple method to set the reference values from the results of some few supercell calculations.
Despite the simplicity of this method, we show that it works quite well for the construction of the CPA Green's function within the Wannier formalism.
\par
In the following, we first present our formulation of the CPA in terms of the Wannier representation.
As examples for its application, we show results for the Bloch spectral function, the density of states (DOS), and the magnetic moment in the Fe-based 3d transition-metal alloys, Fe-V, Fe-Co, Fe-Ni, and Fe-Cu.
We verify the accuracy of the Wannier-CPA method by comparing the results with those calculated via the KKR-CPA method.
Despite the rather simple formulation for the Wannier-CPA method, these quantities obtained by this way reproduce quite well the results obtained by the more demanding KKR-CPA method.
\section{II. Formulation}
To develop a simple and general computational method for homogeneous random alloys, we formulate the CPA on the basis of the Wannier formalism.
We evaluate the performance of the Wannier-CPA method by comparing with results obtained via the well-developed KKR-CPA calculation method.
We first present the formulation of the CPA in random alloys as used within the KKR-CPA method, and then adapt it for the Wannier representation.
\subsection{A. KKR-CPA}
The most prominent feature of the KKR Green's function method is that the Green's function of the system is already set up and used during the SCF calculations.
Within the KKR formalism the Green's function for a pure system is given as follows \cite{Weinberger1990Oxford}:
\begin{align}
  &G({\bm r}+{\bm R}_I+{\bm Q}_i,{\bm r}'+{\bm R}_J+{\bm Q}_j,E) \notag \\
    &\quad
    =
      \sum_{\Lambda,\Lambda'}
      Z^i_\Lambda({\bm r},E)
      \tau^{IiJj}_{\Lambda\Lambda'}(E)
      Z^{j\times}_{\Lambda'}({\bm r}',E) \notag \\
    &\quad\qquad
      -\delta_{IJ}\delta_{ij}
      \sum_{\Lambda}
      \big[
        Z^i_\Lambda({\bm r},E)J^{i\times}_\Lambda({\bm r}',E)\theta(r'-r) \notag \\
    &\quad\qquad
        +J^i_\Lambda({\bm r},E)Z^{i\times}_\Lambda({\bm r}',E)\theta(r-r')
      \big]\;.
\end{align}
Here, ${\bm R}_I$ and ${\bm Q}_i$ give the position of the unit cell $I$ and atomic site $i$, respectively, and ${\bm r}$ and ${\bm r}'$ refers to the position of electrons on atoms siting at $({\bm R}_I,{\bm Q}_i)$ and $({\bm R}_J,{\bm Q}_j)$, respectively.
The functions $Z^i_\Lambda$ and $J^i_\Lambda$ stand for the regular and irregular solutions of the single-site Schr\"odinger or Dirac equation for site $i$, respectively.
In the relativistic formulation, the subscript $\Lambda=(\kappa,\mu)$ stands for the combination of the relativistic spin-orbit ($\kappa$) and magnetic ($\mu$) quantum numbers and the superscript of $\times$ refers to the left-hand side solution of the Dirac equation \cite{Ebert2011RepProgPhys}.
The general definition of the scattering path operator $\tau^{IiJj}_{\Lambda\Lambda'}(E)$ accounts for all scattering events connecting site $i$ of $I$-th unit cell and site $j$ of $J$-th unit cell.
\par
As the Green's function of the system is obtained directly when using the KKR method, it is easy to calculate physical quantities or incorporate the effect of alloying by means of the CPA, since these can be formulated in terms of the Green's function.
The CPA condition of the KKR formalism is formulated by the following equations \cite{Gyorffy1972PRB}:
\begin{align}
  \label{eq:KKR-CPA}
  \underline{\tau}^{0i0i}_c(E)
    &=
      \sum_\alpha
      c_\alpha
      \underline{\tau}^{0i0i}_\alpha(E)\;, \\
  \underline{\tau}^{0i0i}_c(E)
    &=
      \frac{1}{\Omega_{\rm BZ}}
      \int_{\rm BZ}{\rm d}^3k\;
      \left[
        {\underline{\underline{t}}{}_c}^{-1}(E)
        -\underline{\underline{\mathcal{G}}}{}_0({\bm k},E)
      \right]_{ii}^{-1}\;, \\
  \underline{\tau}^{0i0i}_\alpha(E)
    &=
      \left[
        {\underline{t}^i_\alpha}^{-1}(E)
        -{\underline{t}^i_c}^{-1}(E)
        +{\underline{\tau}^{0i0i}_c}^{-1}(E)
      \right]^{-1}\;,
\end{align}
where $\underline{\tau}^{0i0i}_c(E)$ is the site-diagonal CPA scattering path operator, the subscript $\alpha$ is the index for the atom types in the alloy, $c_\alpha$ is the concentration of atom $\alpha$, and the underline indicates matrices with respect to the combined spin-angular momentum index $\Lambda$.
The information on the coherent potential in a random alloy is contained in $\underline{t}^i_c(E)$ which is the single-site scattering matrix for the coherent potential.
For ordered systems, the CPA scattering path operator is given by a Brillouin zone integral in terms of the CPA single-site scattering matrix and so-called KKR structure constant $\underline{\mathcal{G}}^{ij}_0({\bm k},E)$.
The final equation gives the scattering path operator for an embedded $\alpha$-atom on the site $i$ into the CPA medium.
\par
Unlike the mentioned VCA and ATA methods, we have to determine the CPA medium self-consistently for the given $t$-matrices $\underline{t}^i_\alpha(E)$ of the components.
Several algorithms have been suggested to deal with the above CPA equations.
The most commonly used algorithm was worked out by Mills et al. \cite{Mills1983PRB}, and allows to obtain the CPA scattering path operator $\underline{\tau}^{0i0i}_c(E)$ by an iterative process (for more details see Ref. \cite{Ginatempo1988JPhysFMetPhys}).
\subsection{B. Wannier-CPA}
While the Green's function is directly supplied by the KKR-CPA method, we first have to construct a TB Hamiltonian in case of the Wannier formalism according to the expression:
\begin{align}
  \mathcal{H}
    =
      \sum_{I,J}\sum_{i,j}\sum_{n,n'}
      \ket{{\bm R}_I+{\bm Q}_i,n}H^{IiJj}_{nn'}
      \bra{{\bm R}_J+{\bm Q}_j,n'}\;,
\end{align}
where $n$ is the index of the Wannier functions including spin.
To make use of the CPA, we divide the Hamiltonian into site diagonal and off-diagonal terms as in CPA, in which a single-site theory is formulated only for diagonal terms:
\begin{align}
  H^{IiJj}_{nn'}
    =
      (1-\delta_{IJ}\delta_{ij})
      t^{IiJj}_{nn'}
      +\delta_{IJ}\delta_{ij}
      v^i_{nn'}\;,
\end{align}
where $v$ and $t$ are the on-site potential and the site off-diagonal terms of the Hamiltonian of the Wannier basis, respectively.
To apply the CPA to the Wannier functions, we construct the Green's function of the system from the TB Hamiltonian.
Within the Wannier representation, the corresponding real-space Green's function is given by an integral over the Brillouin zone (BZ) as follows:
\begin{align}
  G^{0iJj}_{nn'}(E)
    =
      \frac{1}{\Omega_{\rm BZ}}
      \int_{\rm BZ}{\rm d}^3k\;
      \mathcal{G}^{ij}_{nn'}({\bm k},E)
      e^{-i{\bm k}\cdot{\bm R}_J}\;,
\end{align}
where $\Omega_{\rm BZ}$ is the volume of the BZ, and $\mathcal{G}^{ij}_{nn'}({\bm k},E)$ is the Fourier transform of the Green's function.
Here, the Fourier transform is used to reduce the computation time by reducing the number of matrix elements in the Green's function when compared to the real space representation.
In matrix form, $\mathcal{G}^{ij}_{nn'}({\bm k},E)$ satisfies the following equation:
\begin{align}
  \underline{\underline{\mathcal{G}}}({\bm k},E)
    =
      \left[
        \underline{\underline{\mathcal{G}}}{}_0^{-1}(E)
        -\underline{\underline{\mathcal{T}}}({\bm k})
      \right]^{-1}\;,
\end{align}
where matrices with both atomic site and the Wannier function indices are indicated by a double-underline.
Here, the matrix elements are given by
\begin{align}
  \left[\underline{\underline{\mathcal{G}}}({\bm k},E)\right]^{ij}_{nn'}
    &=
      \mathcal{G}^{ij}_{nn'}({\bm k},E)\;, \\
  \left[\underline{\underline{\mathcal{G}}}{}_0^{-1}(E)\right]^{ij}_{nn'}
    &=
      \delta_{ij}
      (\delta_{nn'}E-v^i_{nn'})\;, \\
  \left[\underline{\underline{\mathcal{T}}}({\bm k})\right]^{ij}_{nn'}
    &=
      \sum_J(1-\delta_{0J}\delta_{ij})
      t^{0iJj}_{nn'}
      e^{i{\bm k}\cdot{\bm R}_J}\;.
\end{align}
\par
The KKR multiple scattering formulation for the CPA condition cannot be used within the Wannier formalism.
This is because of the complexity in defining the scattering path operator in the TB model.
Therefore, we used an equation mathematically equivalent to the first equation in the KKR-CPA condition (see Eq. (\ref{eq:KKR-CPA})).
For this purpose, we exploit the representation of the scattering operator $\underline{t}^i_\alpha(E)$,
\begin{align}
  \underline{t}^i_\alpha(E)
    =
      \left(
        \underline{v}^i_\alpha
        -\underline{v}^i_c(E)
      \right)
      \left[
        1
        -\underline{G}^{0i0i}_c(E)
        \left(
          \underline{v}^i_\alpha
          -\underline{v}^i_c(E)
        \right)
      \right]^{-1}\;,
\end{align}
in which the fictitious coherent potential $\underline{v}^i_c(E)$ is replaced by the real potential of $\alpha$-atom $\underline{v}^i_\alpha$ at site $i$.
Herein, $\underline{G}^{0i0i}_c(E)$ corresponds to the Green's function of the coherent potential in the reference unit cell.
The CPA condition in the single-site approximation is then given by
\begin{align}
  \label{eq:Wannier-CPA}
  \braket{\underline{t}^i}
    =
      \sum_\alpha
      c_\alpha
      \underline{t}^i_\alpha
    =
      0\;,
\end{align}
where we indicate the matrices with respect to the combined indices of Wannier functions $n$ by single-underline.
For the numerical calculation of the coherent potential, $\underline{v}^i_c(E)$, we use an iterative method that is similar to the Mills' algorithm.
We update the $n$-th temporary coherent potential ${\underline{v}^i_c}^{(n)}(E)$ in the following way.
When the CPA condition is not satisfied by the $n$-th temporary coherent potential, we can define the concentration averaged scattering operator as
\begin{align}
  \braket{\underline{t}^i}^{(n)}
    =
      \sum_\alpha
      c_\alpha
      {\underline{t}^i_\alpha}^{(n)}
    \neq
      0\;.
\end{align}
The next update for the coherent potential is obtained as follows:
\begin{align}
  {\underline{v}^i_c}^{(n+1)}(E)
    =
      {\underline{v}^i_c}^{(n)}(E)
      +\braket{\underline{t}^i}^{(n)}
      \left(
        1
        +{\underline{G}^{0i0i}_c}^{(n)}(E)
        \braket{\underline{t}^i}^{(n)}
      \right)^{-1}\;,
\end{align}
where the Green's function is obtained from the $n$-th coherent potential ${\underline{v}^i_c}^{(n)}(E)$.
We repeat the cycle until $\braket{\underline{t}_i}^{(n)}$ becomes smaller than a threshold $\delta$.
We exploit the VCA for the initial guess of the coherent potential as follows:
\begin{align}
  {\underline{v}^i_c}^{(1)}(E)
    =
      \sum_\alpha
      c_\alpha
      \underline{v}^{i}_\alpha\;.
\end{align}
\par
To apply this formalism to real alloys, we have to consider the following two points.
One is the on-site potentials of the two pure components, since the DFT-based Wannier Hamiltonian does not give information on the reference value of these potentials.
To determine the relative on-site potential energies, we use the supercell calculations as follows.
Let us consider an $A$-$B$ binary alloy.
First, we perform DFT calculations for a supercell of eight-atoms, $A_1B_7$ and $A_7B_1$, and construct the TB Hamiltonian in the Wannier basis.
Then, we derive the difference of the on-site potential of the $3d$ orbitals between component $A$ and $B$ in both $A_1B_7$ and $A_7B_1$, and set its average as $\Delta v_{A-B}^{\rm supercell}$.
Then, we perform DFT calculations for pure $A$ and pure $B$, and construct the Wannier TB Hamiltonian.
We use this Hamiltonian for the calculation of the CPA Green's function, but before starting the CPA calculation, we subtract a constant from the diagonal terms of the on-site potential so that the potential difference of $3d$ orbitals in pure $A$ and pure $B$ becomes $\Delta v_{A-B}^{\rm supercell}$.
The other point to consider is the determination of the site off-diagonal term in the TB Hamiltonian since the site diagonal terms as well as the site off-diagonal terms are different for the two components.
This is in sharp contrast to the KKR-CPA formalism, where only the scattering path operator depends on the component.
In this paper, since we consider the alloys consisting of two transition metal elements, we simply take a concentration average \cite{Gyorffy1974JPhysColloq,Richter1987JPhysFMetPhys}.
\par
The accurate determination of the Fermi energy is important for examining the magnetic properties of alloys.
We set the Fermi energy so that the total number of electrons
\begin{align}
  N
    =
      -\frac{1}{\pi}
      {\rm Im}
      {\rm Tr}
      \int^{E_{\rm F}}{\rm d}E
      \underline{\underline{G}}{}_c(E)\;,
\end{align}
is consistent with the number of electrons in the $A$-$B$ alloy.
A complex contour Gauss-Legendre integral is used to calculate the above integral.
The Fermi energy of the system is determined iteratively using the DOS and the difference between the total number of electrons of the alloy and the number of electrons obtained by integrating the Green's function up to the temporary Fermi energy.
\par
\subsection{C. Computational Steps of a Wannier-CPA Calculation}
The CPA calculations using the Wannier formalism have been organized as follows:
First, we performed DFT calculations using the QUANTUM ESPRESSO package \cite{Giannozzi2009JPhysCondensMatter,Giannozzi2017JPhysCondensMatter} based on plane waves and pseudopotentials.
We use the ultrasoft pseudopotentials \cite{DalCorso2005PRB} in the PSlibrary \cite{Corso2014ComputMaterSci} with the functional type of GGA-PBE exchange-correlation functional \cite{Perdew1996PRL} and with relativistic effects included.
Here, we set the lattice constant as the experimental value of bcc Fe $a=2.86$ \AA\; assuming that bcc Fe is alloyed with other transition-metal elements.
\par
The wannierization process is conducted by using WANNIER90 package \cite{Marzari1997PRB,Souza2001PRB,Mostofi2008ComputPhysCommun,Mostofi2014ComputPhysCommun,Pizzi2020JPhysCondensMatter} to reproduce the DFT energy bands below $E_{\rm F}$ + 3 eV, with $E_{\rm F}$ being the Fermi energy.
We construct for each spin a nine-orbital model, which contains one 4s, five 3d, and three 4p atomic orbitals.
As the relative values of the reference for the on-site potential is not given by the Wannier Hamiltonian, we determined the difference of the on-site potential of $3d$ orbitals between Fe and $X$ ($X$ = V, Co, Ni, and Cu) components by using $\Delta v_{{\rm Fe}-X}^{\rm supercell}$ calculated in nonmagnetic mode.
Using the on-site potential, we perfumed the Wannier-CPA calculations.
We will discuss the dependence of the results on the constant subtracted from the on-site potential of $X$ in magnetic moment in the last part of the next section.
\par
The electric structure calculation of KKR-CPA method is performed self-consistently by the fully relativistic spin-polarized Munich SPR-KKR package \cite{Ebert2011RepProgPhys,EbertSPRKKR}.
For the exchange-correlation functional, the parametrization given by Vosko {\it et al.} \cite{Vosko1980CanJPhys} has been used.
An angular momentum cutoff of $l_{\rm max}=4$ was used for the KKR multiple-scattering calculations.
Here, we used the same lattice parameter $a=2.86$ \AA\; as in the Wannier-CPA calculation.
\par
As the Green's function of a random alloy is obtained by the process described above, we can calculate the Bloch spectral function, the DOS, and the magnetic moment from the obtained CPA Green's function.
In the following section, we describe results of calculations of these quantities in transition-metal alloys and compare them with the results obtained by the KKR-CPA method.
\section{III. Results and Discussion}
In this section, we present results for various physical quantities obtained by the Wannier-CPA method. We discuss their accuracy by comparing the Bloch spectral function, the DOS, and the magnetic moment calculated by the Wannier-CPA and the KKR-CPA methods, respectively.
As an interesting target material, we focus on the Fe-based 3d transition-metal alloys Fe-$X$ ($X$ = V, Co, Ni, and Cu).
We selected the elements $X$ so that the atomic number of $X$ is close to that of Fe.
Here, we omit Fe-Cr and Fe-Mn alloys, for which an antiferromagnetic configuration is predicted in the alloy systems \cite{Hirai1998JPSJ,Sakuma1998JPSJ}.
As is demonstrated in the following, we found an excellent agreement between the results obtained by the Wannier-CPA and the KKR-CPA methods for the considered alloys.
\subsection{A. Bloch Spectral Function}
\begin{figure}[t]
\begin{center}
\includegraphics[width=8.5cm]{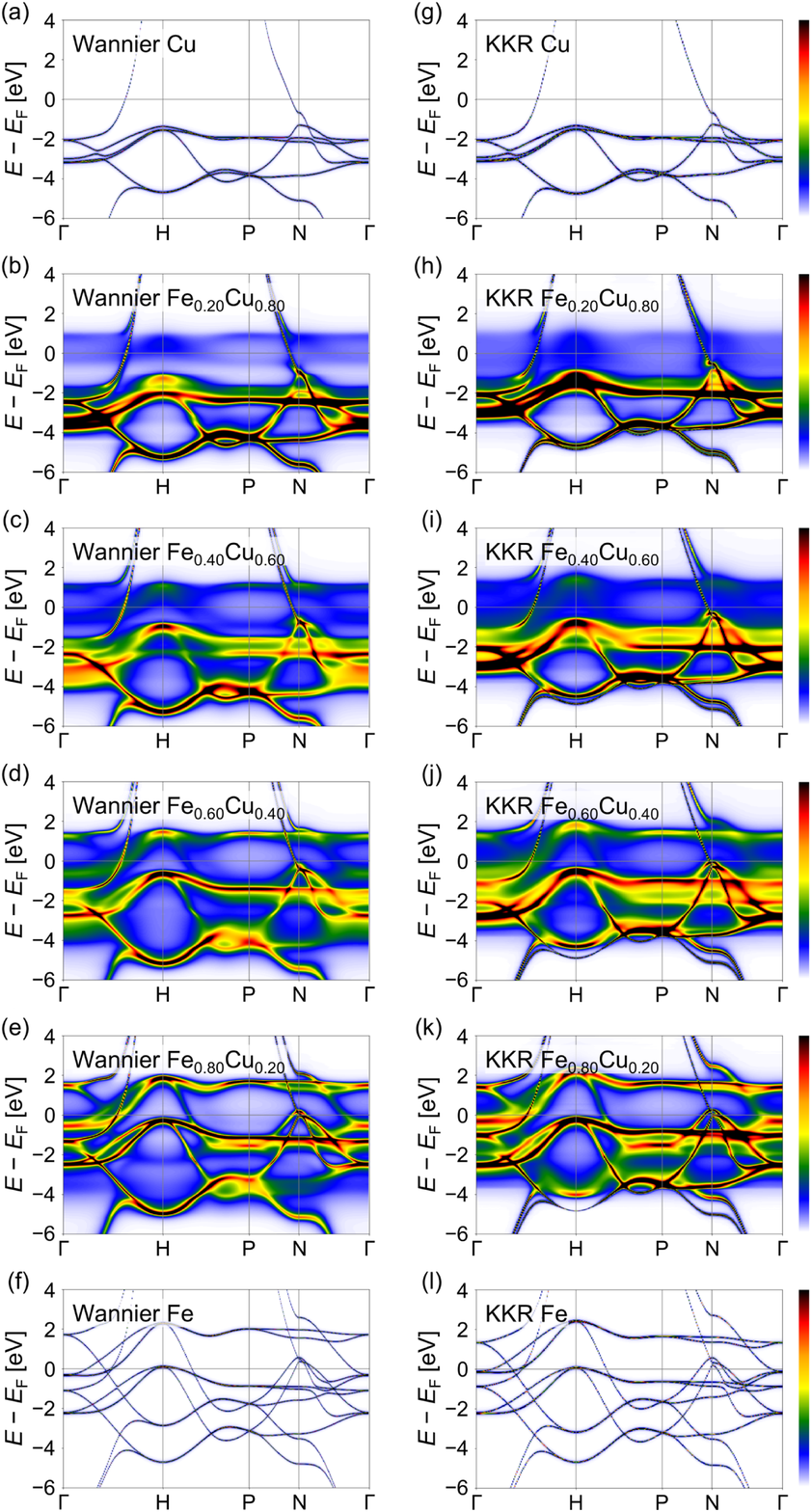}
\end{center}
\caption{Bloch spectral functions along the high symmetry lines $\Gamma$-H-P-N-$\Gamma$ for Fe$_x$Cu$_{1-x}$ with the concentrations (a) $x=0.00$, (b) $x=0.20$, (c) $x=0.40$, (d) $x=0.60$, (e) $x=0.80$, and (f) $x=1.00$ calculated by the Wannier-CPA method, and (g) $x=0.00$, (h) $x=0.20$, (i) $x=0.40$, (j) $x=0.60$, (k) $x=0.80$, and (l) $x=1.00$ calculated by the KKR-CPA method.}
\label{fig:Bloch_all}
\end{figure}

First, we show the Bloch spectral functions of bcc Fe-Cu alloys calculated by both the Wannier-CPA and the KKR-CPA methods to compare the basic electric structure that determines physical quantities.
The Bloch spectral function is the imaginary part of the trace of the Green's function given as follows:
\begin{align}
  A({\bm k},E)
    =
      -\frac{1}{\pi}
      {\rm Im}\;
      {\rm Tr}\;
      \underline{\underline{\mathcal{G}}}{}_c({\bm k},E)\;.
\end{align}
If we plot the wave vector and energy region where the Bloch spectral function takes finite values, it shows a structure very similar to the band structure or dispersion relation $E({\bm k})$ of the pure systems.
Figure \ref{fig:Bloch_all} shows the representative Bloch spectral function for bcc Fe$_x$Cu$_{1-x}$ ($x$ = 0.0, 0.2, 0.4, 0.6, 0.8, and 1.0) alloys calculated by both the Wannier-CPA (left side) and the KKR-CPA (right side) methods.
We obtained a similar behavior for the Bloch spectral function of Fe-V, Fe-Co, and Fe-Ni alloys.
For this reason, we discuss only the details on the calculations of the bcc Fe-Cu alloys.
The calculations were performed by using the bcc structure for all the calculations on Fe-Cu alloys for simplicity, although Cu takes fcc structure in the pure form.
Similarly, the calculations of Fe-V, Fe-Co, and Fe-Ni alloys were also performed by using the bcc structure.
For the calculation for pure Fe and Cu, we added a small imaginary part of 0.1 mRy to the energy to obtain visible Bloch spectra because the spectral structure consists of a delta function for pure Fe and Cu.
As shown in Fig.~\ref{fig:Bloch_all}, we obtain very close spectral structures from the Wannier-CPA and the KKR-CPA methods.

\begin{figure}[t]
\begin{center}
\includegraphics[width=8.5cm]{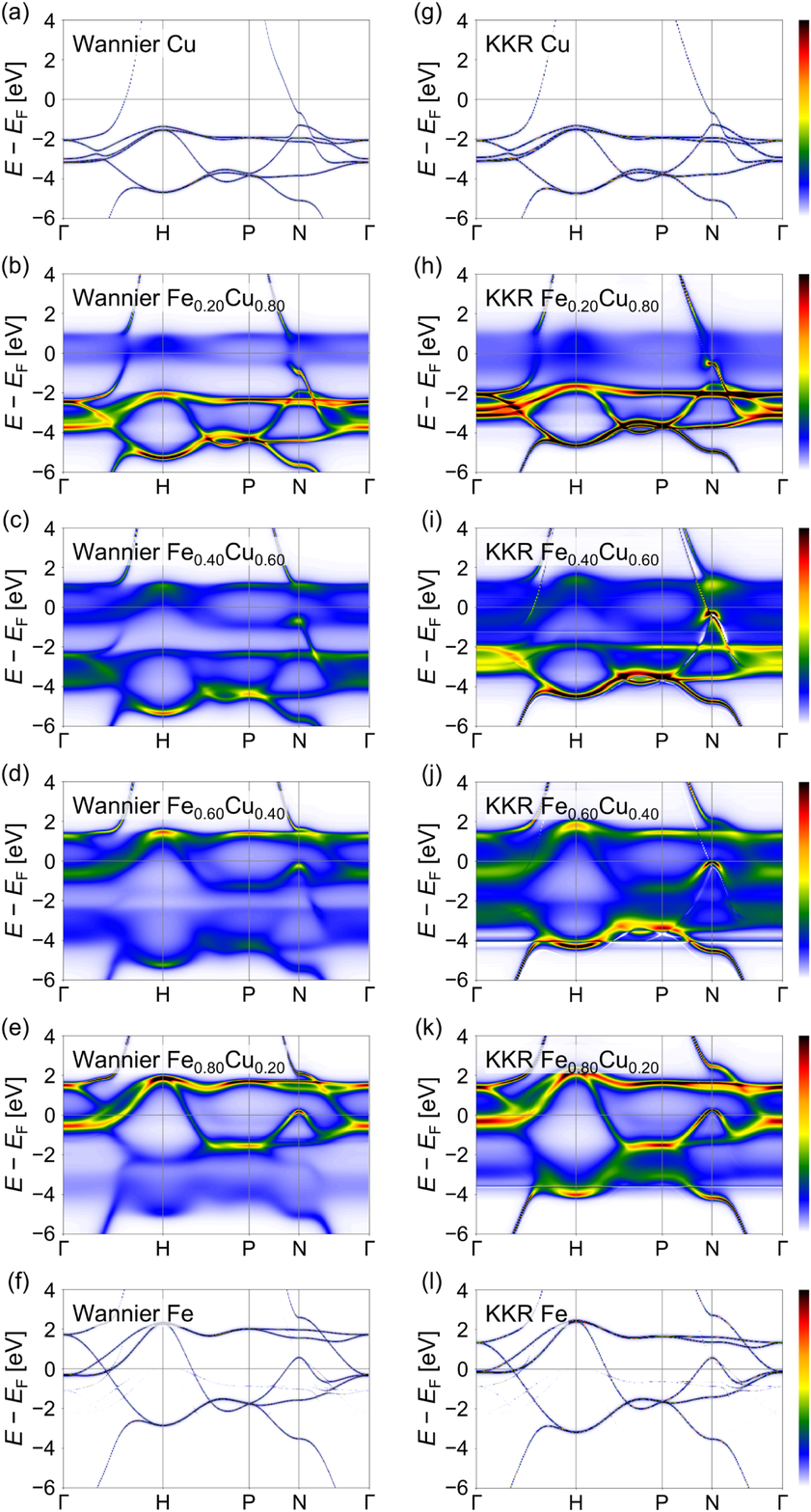}
\end{center}
\caption{As for Fig.~\ref{fig:Bloch_all} but projected to spin-down states.}
\label{fig:Bloch_down}
\end{figure}

\begin{figure}[t]
\begin{center}
\includegraphics[width=8.5cm]{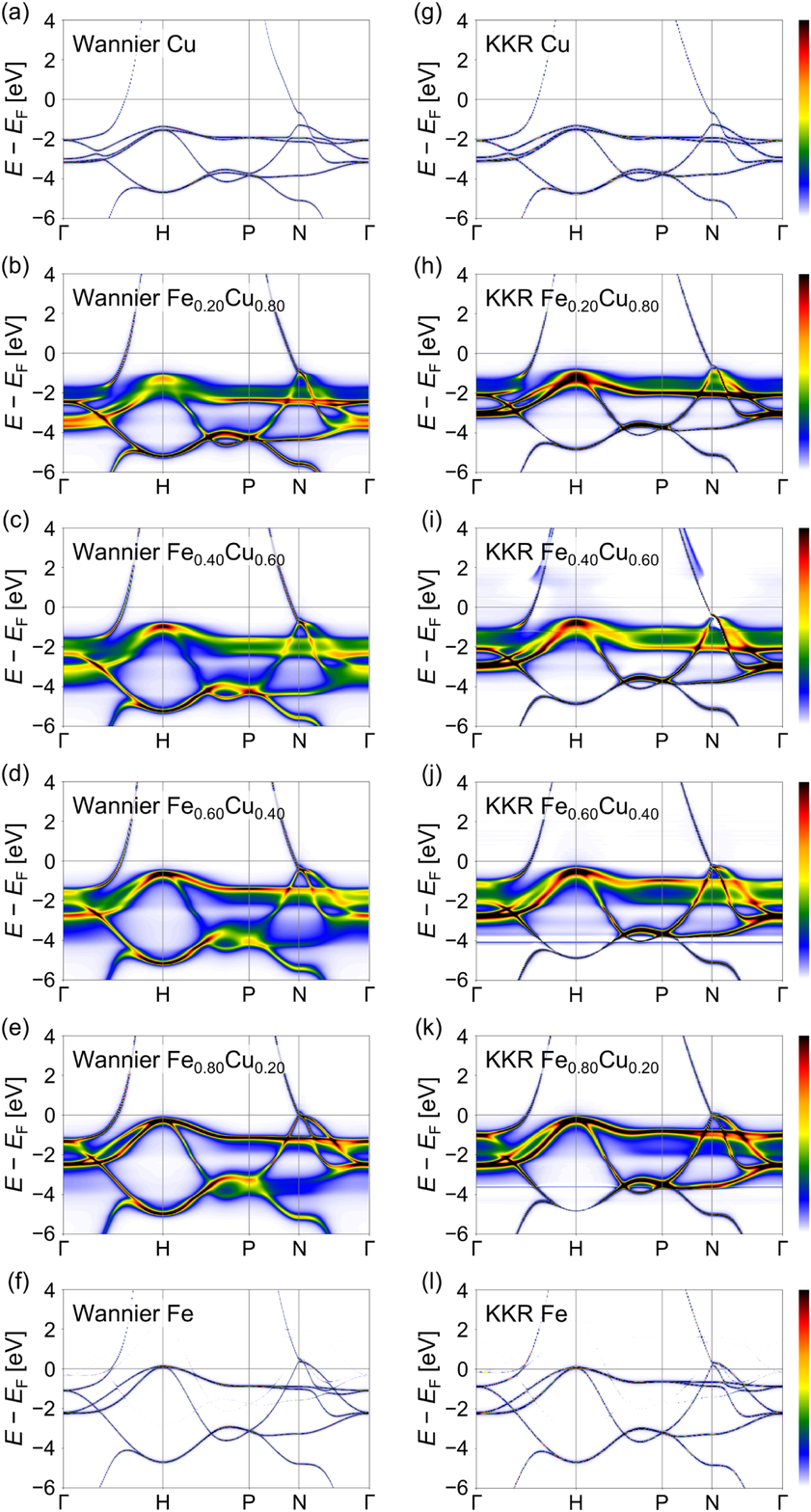}
\end{center}
\caption{As for Fig.~\ref{fig:Bloch_all} but projected to spin-up states.}
\label{fig:Bloch_up}
\end{figure}

Since the Bloch spectral functions of the total state have a rather complex structure, we resolved them with respect to the spin directions.
The corresponding Bloch spectral functions for the spin-down- and up-states in Fe$_x$Cu$_{1-x}$ are represented in Figs.~\ref{fig:Bloch_down} and \ref{fig:Bloch_up}, respectively.
As the concentration is changed from Cu to Fe$_{0.20}$Cu$_{0.80}$, the spectral structure of Fe slightly appears between $-$1 eV and 1 eV in both the Wannier-CPA and the KKR-CPA results as shown in Figs.~\ref{fig:Bloch_down}(b) and \ref{fig:Bloch_down}(h), which is about 2 eV lower than that for the pure Fe.
This spectral structure becomes clear and shifts to higher energies as the concentration of Fe is increased to 0.40 and 0.60 in both the methods.
\par
The only major difference of the Bloch spectral function between the Wannier-CPA and the KKR-CPA results appears in a spectral structure of Fe$_{0.80}$Cu$_{0.20}$ near the H-point in the reciprocal lattice.
We observed a pronounced structure near $-$2 eV in the case of the Wannier-CPA method as shown in Fig.~\ref{fig:Bloch_down}(e), which is strongly affected by the spectral structure of Fe.
On the other hand, in the KKR-CPA method, this feature is mixed with the spectrum of Cu at around $-$5 eV forming a single peak structure [Fig.~\ref{fig:Bloch_down}(k)].
\par
For pure bcc Fe, we observed weak spectral features in the spin-down channel that reflect the main spectra of the spin-up states between $-$2 eV and 2 eV in the results obtained by the Wannier-CPA and the KKR-CPA methods [Figs.~\ref{fig:Bloch_down}(f) and \ref{fig:Bloch_down}(l)].
These weak features can be ascribed to the relativistic effect of the mixing of spin-up and spin-down states by the spin-orbit coupling.

Figure \ref{fig:Bloch_up} shows the Bloch spectral functions for the spin-up states in Fe$_x$Cu$_{1-x}$ calculated by both the Wannier-CPA and the KKR-CPA methods.
Unlike the spectral structure of Fe$_{0.20}$Cu$_{0.80}$ in the spin-down state, the blurred spectral structure near the Fermi energy does not show up in the spin-up state as shown in Figs.~\ref{fig:Bloch_up}(b) and \ref{fig:Bloch_up}(h), forming a sharper structure over the entire region.
This so called virtual crystal like behavior indicates that the spin-up spectra of pure Fe and Cu are energetically closer to each other than those of the spin-down states.
For the same reason, the spectral structures between $-$6 eV and $-$4 eV are less blurred compared with those between $-$3 eV and $-$1 eV in Fe$_x$Cu$_{1-x}$ ($x=0.20\sim0.80$) alloys since the spectral structure of the pure Fe and Cr are energetically closer to each other between $-$6 eV and $-$4 eV.
This behavior was observed for both the Wannier-CPA and the KKR-CPA methods [Figs.~\ref{fig:Bloch_up}(b)-(e) and \ref{fig:Bloch_up}(h)-(k)].
We again observed weak spectral feature corresponding to the main spectral structure of the spin-down states in the pure Fe between the energy of $-$2 eV and 1 eV [Figs.~\ref{fig:Bloch_up}(f) and \ref{fig:Bloch_up}(l)].

\subsection{B. Density of States}
\begin{figure}[t]
\begin{center}
\includegraphics[width=8.5cm]{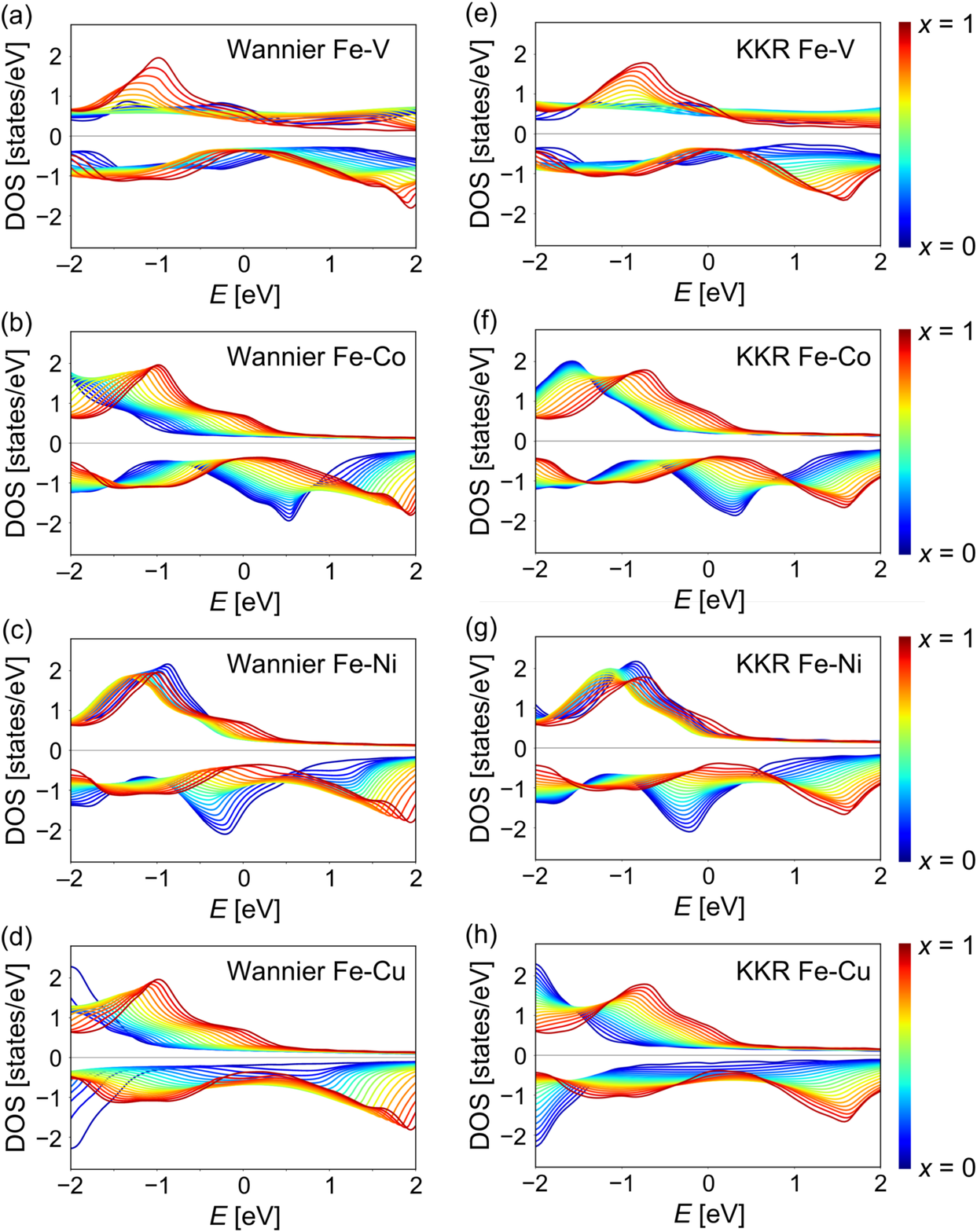}
\end{center}
\caption{DOS of Fe$_xX_{1-x}$ where $X$ atoms are (a) V, (b) Co, (c) Ni, and (d) Cu alloys calculated by the Wannier-CPA method, and (e) V, (f) Co, (g) Ni, and (h) Cu alloys calculated by the KKR-CPA method. The DOS of pure Fe is given by red lines, and that of $X$ is given by blue lines. For alloys, we give the DOS by intermediate colors between red and blue depending on the concentration of Fe and $X$.}
\label{fig:DOS}
\end{figure}

Figure \ref{fig:DOS} shows the computational results for the DOS near the Fermi energy ($-$2 eV to 2 eV) obtained by the Wannier-CPA and the KKR-CPA methods for bcc Fe-$X$ ($X$ = V, Co, Ni, and Cu) alloys to monitor the occupation trend of the states in each alloy.
Here, the DOS is given by integrating the Bloch spectral functions over the BZ:
\begin{align}
  D(E)
    =
      \int_{\rm BZ}{\rm d}^3k\;
      A({\bm k},E)\;.
\end{align}
We plotted the DOS of pure Fe and the pure $X$ component with red and blue lines, respectively.
For the alloy systems, we plotted the DOS with a neutral color between red and blue depending on the concentration of Fe and $X$.
Figure \ref{fig:DOS} shows that the qualitative behavior of the DOS energy shift with increasing $X$ concentration is fully consistent for both the Wannier-CPA and the KKR-CPA methods.
A representative example can be seen in the up-DOS of Fe-Ni alloys.
For the KKR-CPA method [Fig.~\ref{fig:DOS}(g)], the peak structure arising from Fe at around $-$1 eV is shifted to lower energies as the concentration of Ni increases, taking a minimum at around Fe$_{0.50}$Ni$_{0.50}$, and is then shifted to higher energies.
This behavior is reproduced quite well by the Wannier-CPA calculations as shown in Fig.~\ref{fig:DOS}(c).

\begin{figure}[t]
\begin{center}
\includegraphics[width=8.5cm]{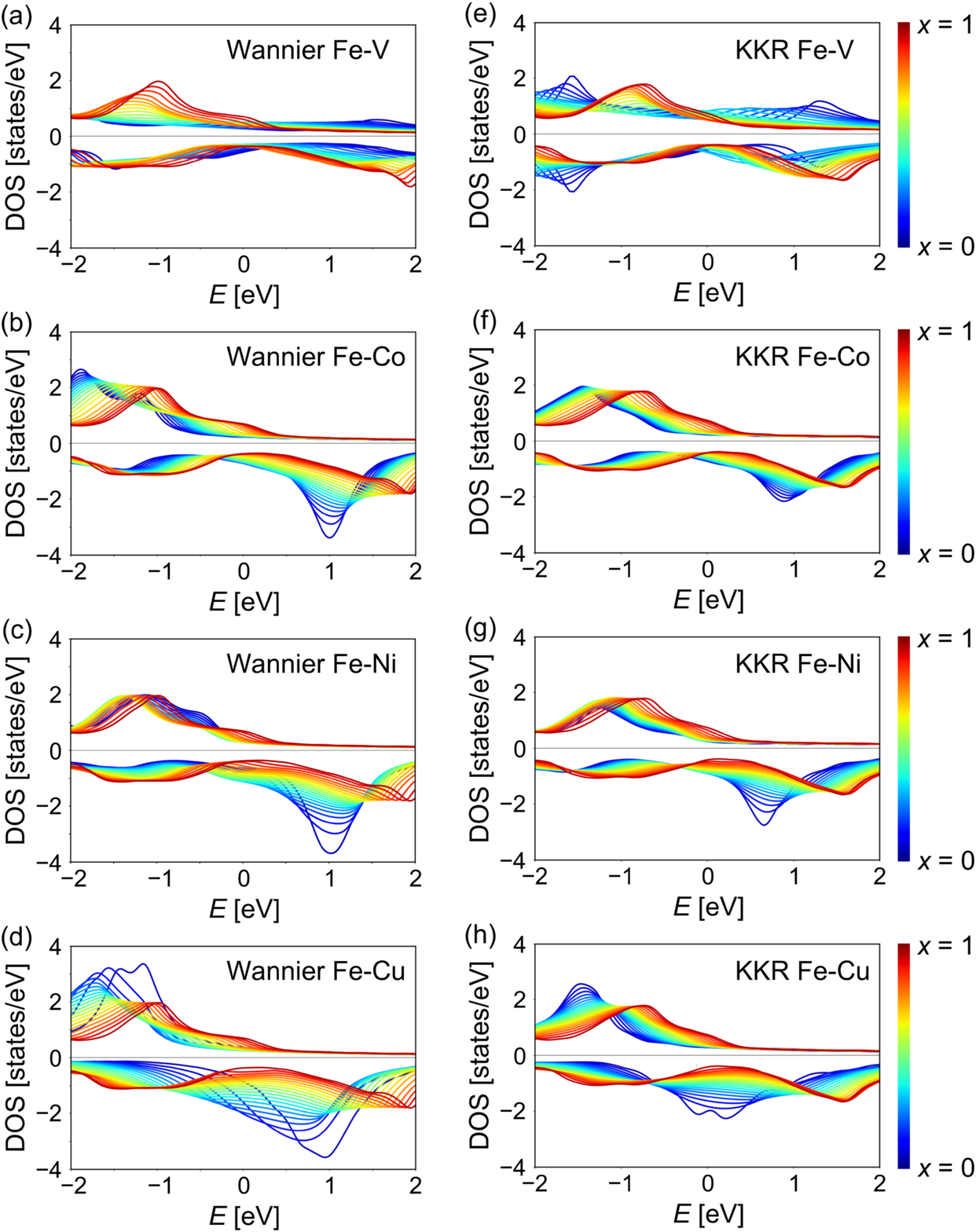}
\end{center}
\caption{As for Fig.~\ref{fig:DOS} but projected to Fe-component.}
\label{fig:DOS_Fe}
\end{figure}

\begin{figure}[t]
\begin{center}
\includegraphics[width=8.5cm]{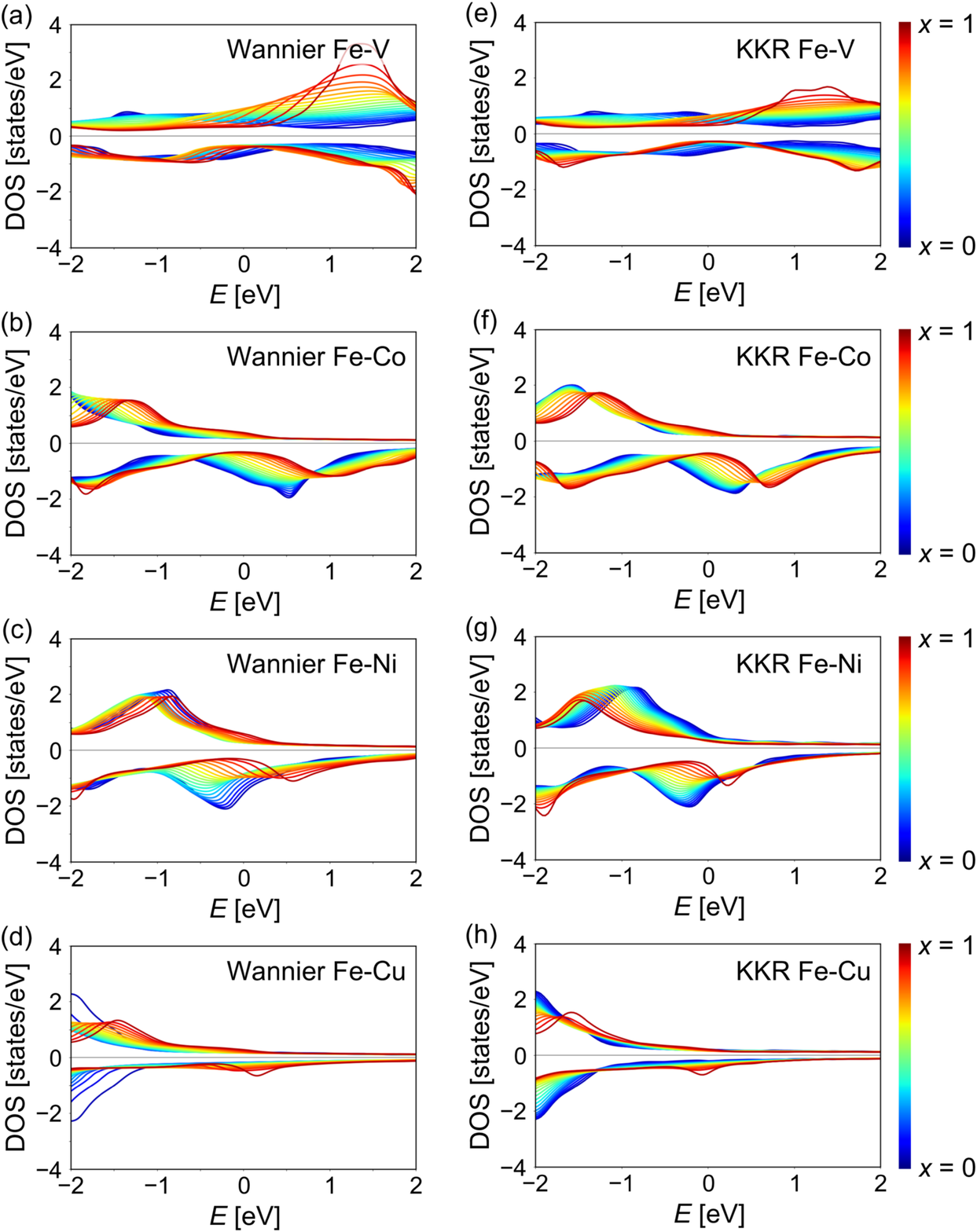}
\end{center}
\caption{As for Fig.~\ref{fig:DOS} but projected to $X$-component ($X$ = V, Co, Ni, and Cu).}
\label{fig:DOS_X}
\end{figure}

To examine the element-specific properties of the DOS, we define the component projection of the Green's function.
According to Ref. \cite{Gyorffy1972PRB}, the CPA condition given by Eq. (\ref{eq:Wannier-CPA}) can be rewritten using the Green's function as follows:
\begin{align}
  \underline{G}{}^{0i0i}_c(E)
    =
      \sum_\alpha
      c_\alpha
      \underline{G}{}^{0i0i}_\alpha(E)\;,
\end{align}
where $\underline{G}{}^{0i0i}_\alpha(E)$ is given by
\begin{align}
  \underline{G}{}^{0i0i}_\alpha(E)
    =
      \underline{G}{}^{0i0i}_c(E)
      +\underline{G}{}^{0i0i}_c(E)
      \underline{t}{}^i_\alpha(E)
      \underline{G}{}^{0i0i}_c(E)\;.
\end{align}
Here, $\underline{G}{}^{0i0i}_\alpha(E)$ gives the Green's function when for the site $i$ of the $0$-th unit cell the $t$-matrix of the CPA medium is replaced by that for component $\alpha$.
Therefore, $\underline{G}{}^{0i0i}_\alpha(E)$ corresponds to the $\alpha$-component projection of the CPA Green's function.
As the component projection of the Green's function is determined in the CPA cycle, the calculation of the element-specific properties of the DOS is straightforward.
Figures \ref{fig:DOS_Fe} and \ref{fig:DOS_X} show the Fe and $X$-specific DOS of Fe-$X$ ($X$ = V, Co, Ni, and Cu) alloys, respectively.
Here, again, the DOS of the alloy with a high Fe concentration is plotted with reddish lines, and that with a high $X$ concentration is given with bluish lines.
On the whole, structural similarities in the element specific DOS calculated by both the Wannier-CPA and the KKR-CPA methods can be found in these figures, but we can also see some small differences in the detailed structure.
For example, the Fe-component projection of the DOS of Fe-V alloys in the V-rich region obtained by KKR-CPA method has a peak structure near $-$1.5 eV [Fig.~\ref{fig:DOS_Fe}(e)], which are not observed in the up-DOS and quite small in the down-DOS of the Wannier-CPA calculation [Fig.~\ref{fig:DOS_Fe}(a)].

\subsection{C. Magnetic Moment}
In the previous two subsections, we found that the Wannier-CPA method can reproduce the Bloch spectral function and the DOS of Fe-based transition-metal alloys quite well.
Finally, we discuss the predicted physical quantities by the Wannier-CPA method.
As an example, we focused on the magnetic moment.
Concerning the magnetic moment in Fe-based transition-metal alloys, one of the best benchmarks is the Slater-Pauling curve \cite{Bozorth1951VanNostrand}.
The Slater-Pauling curve is a convex curve that appears when the saturation magnetization of these alloys is plotted against the number of electrons per atom.
In Fe-Co alloys, it is known that the maximum of the saturation magnetization occurs near Fe$_{0.7}$Co$_{0.3}$.
The left and right sides of the curve form a straight line with an angle of 45 degrees with the horizontal axis of the Fe-based alloys when the scale of one electron on the horizontal axis and one Bohr magneton on the vertical axis are equal.
Previous research shows that the experimental results of the Slater-Pauling curve are excellently reproduced by the KKR-CPA calculations \cite{Dederichs1991JMagnMagnMater,Akai1992HyperfineInteract}.
Here, we compare our results for the magnetic moment obtained by our Wannier-CPA method with those of the KKR-CPA method.
\par
\begin{figure}[t]
\begin{center}
\includegraphics[width=9cm]{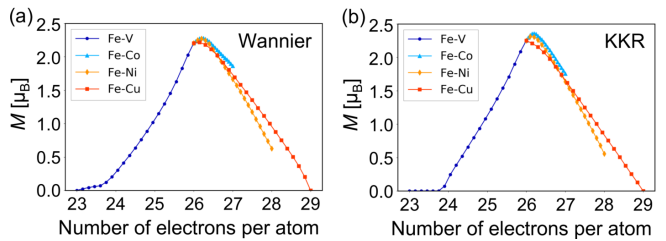}
\end{center}
\caption{Magnetic moment of Fe-$X$ ($X$ = V, Co, Ni, and Cu) calculated by (a) the Wannier-CPA and (b) the KKR-CPA methods.}
\label{fig:Mag}
\end{figure}

Figure \ref{fig:Mag} shows the magnetic moment in the Fe-$X$ ($X$ = V, Co, Ni, and Cu) alloys calculated by the Wannier-CPA and the KKR-CPA methods. 
There is a structural transition from bcc to fcc in the Fe-based alloys when the number of electrons exceeds about 26.7.
However, all the calculations were done in bcc structure for the one-to-one comparison between the Wannier-CPA and the KKR-CPA methods.
We can conclude from Fig.~\ref{fig:Mag} that the Wannier-CPA calculation gives reliable calculation results concerning the calculation of magnetic moments for the following reasons.
First of all, the magnetic moments calculated by the Wannier-CPA method form a typical Slater-Pauling curve, which takes a maximum moment in the case of Fe$_{0.75}$Co$_{0.25}$ and intersects the horizontal axis at an angle of almost 45 degrees.
Furthermore, the calculated magnetic moments in the bcc Fe-$X$ ($X$ = V, Co, Ni, and Cu) alloys by the Wannier-CPA method are in good agreement with those by the KKR-CPA method, since the average values of the deviation in magnetic moments are only 0.057, 0.064, 0.036, and 0.080 $\mu_{\rm B}$, respectively.
These results represent that the Wannier-CPA method can be a powerful tool for the prediction of physical quantities expressed by the integral up to Fermi energy despite its simple formulation.

Since we set the reference values of the on-site potential by a simple method using supercell calculations, we discuss the effect of the change in the magnetic moment due to deviation from the actual reference values of the on-site potential.
Since the difference of $\Delta v_{{\rm Fe}-X}^{\rm supercell}$ ($X$ = V, Co, Ni, and Cu) obtained from Fe$_1X_7$ and Fe$_7X_1$ is within 1 eV \cite{potref}, we calculated the magnetic moment of Fe$_{0.5}X_{0.5}$ by adding 0.5 eV to the diagonal terms of the on-site potential of $X$ in the Wannier-CPA method.
Then, we calculated the magnetic moment by subtracting 0.5 eV from the diagonal terms of the on-site potential of $X$ and derived the difference between the two moments.
We divided it by the magnetic moment calculated without changing the on-site potential and derived the rates of change in magnetic moment.
These rates of change in the magnetic moment were only 6.55\%, 2.07\%, 0.23\%, and 4.72\% in Fe$_{0.5}$V$_{0.5}$, Fe$_{0.5}$Co$_{0.5}$, Fe$_{0.5}$Ni$_{0.5}$, and Fe$_{0.5}$Cu$_{0.5}$, respectively, even with the large difference of 1 eV in the on-site potential. Since the difference in the reference values of the on-site potentials between Fe and $X$ from the actual values has only little effect on the physical quantities, this method using supercell can be a simple and valuable way for determining the relative difference of the on-site potentials of Fe and $X$.
\section{IV. Conclusion}
We have formulated the CPA in the Wannier representation to develop a calculation method for homogeneous random alloys, which can be readily accessed from any first principles calculation methods.
This Wannier-CPA method significantly reduces the computation time compared with those of the existing methods.
Compared to the KKR-CPA method, this Wannier-CPA method can be expected to reduce the computational time by a factor of ten.
To investigate the performance of this Wannier-CPA method, we have examined the Bloch spectral function, the DOS, and the magnetic moment for various Fe-based transition-metal alloys from the Green's function obtained by the Wannier-CPA method, and compared with the results of the calculation by the well-developed KKR-CPA method.
Regarding the Bloch spectral function, the spectral structures of the Fe-Cu alloys were compared by both the Wannier-CPA and the KKR-CPA methods.
We observed a blurred spectral structure of Fe near the Fermi energy in the spin-down state when the Fe content is low.
On the other hand, we observed a clear virtual crystal-like spectral structure in the spin-up state.
This is because of the similarity in the energy structure of Fe and Cu spin-up states.
These behaviors are the same in both the Wannier-CPA and the KKR-CPA methods.
Furthermore, by changing the concentration of Fe, we also found an energy shift in the peak structure of the DOS.
This is the same for the Wannier-CPA and the KKR-CPA calculations for all of the Fe-$X$ ($X$ = V, Co, Ni, and Cu) alloys.
Finally, we calculated the magnetic moment of the Fe-$X$ alloys.
We can reproduce the well-known Slater-Pauling curve by the Wannier-CPA method that is quite similar to the KKR-CPA method, which confirms the good predictive power for physical quantities for the Wannier-CPA method.
In this paper, we have discussed only the Bloch spectral function, the DOS, and the magnetic moment in the Wannier-CPA method.
Nevertheless, one may conclude that this Wannier-CPA method have great applicability to other physical quantities and also large compound systems, which have many restrictions concerning the calculation time as the main bottleneck.
The transport calculation should be one such example.
Although there are many works on the anomalous and spin Hall effect using the Wannier functions, only the intrinsic contribution of the conductivity is considered in all these works.
Using this formulation we have given, it can be possible to calculate the conductivity including the extrinsic contributions as well.
To evaluate the potential of the developed Wannier-CPA method, we expect further applications of the method to various materials in addition to transition-metal alloys.
\section{Acknowledgements}
This work was supported by JSPS KAKENHI Grants Nos.~19H00650, 19H05825, JP19K14607, 21H01003, 21H04437, and 21H04990, Center for Science and Innovation in Spintronics (CSIS) Tohoku University, JST-PREST(JPMJPR20L7), JST-Mirai Program(JPMJMI20A1), Grant-in-Aid for JSPS Fellows Grants No. 21J10577, and GP-Spin, Tohoku University.
K.N. is supported by JST CREST Grant No.~JPMJCR18T2 and 20H01830.
We are grateful to Siobhan Nishimura for English correction of the manuscript.

\newpage


\begin{thebibliography}{99}
\bibitem{potref} The average and the difference of $\Delta v_{{\rm Fe}-X}^{\rm supercell}$ obtained from Fe$_1X_7$ and Fe$_7X_1$ ($X$ = V, Co, Ni, and Cu) are $-$1.75 eV and 0.824 eV for V, 0.803 eV and 0.0656 eV for Co, 0.885 eV and 0.0599 eV for Ni, and 2.28 eV and 0.530 eV for Cu.
\end{thebibliography}
\end{document}